\documentstyle[11pt]{article}
\begin{document}
{\it On the  paper by M. Valluri and D. Merritt 
"Orbital Instability and Relaxation in Stellar Systems"}
\vspace{0.1in}
 
V.G.Gurzadyan
\vspace{0.1in}

Yerevan Physics Institute and University of Rome
\vspace{0.2in}

The paper by M.Valluri and D.Merritt is discussing an interesting and 
difficult topic, and obviously many
questions remain open.  I hope, my comments
will be helpful to understand the reasons of the mentioned "debates". 
 
1. It is not precise to state that "exponential instability of
N-body problem was shown by Miller [1], Gurzadyan and Savvidy [2]..". 
The exponents in [1] and [2] are different quantities and have different
content. 
 
Miller [1] in his pioneering numerical work has shown the exponential
growth of separation of coordinates and velocities for N-body systems. 
It was indeed a remarkable work of early period of computer astronomy.
Due to the development of chaotic dynamics during these decades, 
our present understanding of that problem is much deeper.
 
In [2] we have analytically proved the exponential instability of
spherical
N-body systems in Riemannian space with metric depending on  the
interaction potential, and the positivity of KS-entropy determined by the
curvature (the curvature was zero in Miller's case). This, according to
the
theorems of ergodic theory, means the existence of the property of mixing
and
hence tending to an equilibrium (local). KS-entropy is thus defining the
time
scale of the tending to equilibrium for large N limit, i.e. the
relaxation time; for discussion of the meaning of relaxation time see [3]. 
 
Mixing therefore is not just "a useful way of thinking" but a strictly
defined property of dynamical systems, with sufficient condition
(for typical automorphisms) - the positivity of KS-entropy. 
 
Miller's exponent exists even for N=2, i.e. for two-body problem which is
integrable, with time scale obviously equal to the crossing/orbital time. 
 
While exponent in [2] at N=2 turns to infinity and hence -- no mixing, no
relaxation. 
The systems of 32 particles shown to be exponentially unstable by Miller
[1],
are not unstable via criterion [2]. Indeed, what is the meaning of
relaxation
time for 24 or 32 particles?
 
The time scale obtained numerically by Miller, Valluri and Merritt, as well
as
by other authors, must always be of the order of crossing/dynamical time,
as
actually they had obtained, with no direct relation with relaxation time
scale. Via such numerical experiments hence, for example, nobody had
derived
(and cannot derive), for example, the two-body relaxation time scale,
though
nobody seems to be alarmed by it. 
 
The exponent in [2] describes mixing and hence can be associated with
relaxation, the exponent in [1] - no. 
Thus, not every exponential instability means relaxation (for clear
example
see [4]). The 'debates' quoted by Valluri and Merrit are therefore due to
misunderstanding of this crucial point. 
 
Meanwhile, the relaxation time scale obtained in [2] is supported by: 
 
(a) observational data on globular clusters (see refs in [5]);
 
(b) alternative statistical mechanical methods [6];
 
(c) numerical experiments [7].
 
2. The problem of numerical estimation of Lyapunov numbers for N-body
gravitating systems is itself a non-trivial one, even forgetting that they
are limiting in time quantities. In [8] a theorem is proved showing the
non-equivalence of those Lyapunov numbers and their computer images.
 
3. The massive central object is known to increase the instability and
mixing
properties of N-body systems [9]. 
 
\vspace{0.1in}
 
[1]  R.H.Miller, ApJ, 140, 250, 1964. 
 
[2]  V.G.Gurzadyan, G.K.Savvidy, Doklady AN SSSR,
     277, 69, 1984; AA, 160, 203, 1986.
 
[3]  V.G.Gurzadyan, in: Ergodic Concepts in Stellar Dynamics,
     (Gurzadyan V.G., Pfenniger D., eds.), Springer-Verlag, 1994.
 
[4]  V.G.Gurzadyan, A.A.Kocharyan, AA, 205, 93, 1988. 
 
[5]  G.Meylan (this volume); astro-ph/9912495. 
 
[6]  M.Antoni, S.Ruffo, A.Torcini (this volume); cond-mat/9908336. 
 
[7]  A.A.El-Zant, AA, 326, 113, 1997. 
 
[8]  V.G.Gurzadyan, A.A.Kocharyan, J.Phys.A, 27, 2879, 1994. 
 
[9]  V.G.Gurzadyan, A.A.Kocharyan, Doklady AN SSSR,301, 323, 1988;
     V.G.Gurzadyan, S.G.Matinian, A.A.Kocharyan, Doklady AN SSSR, 296, 54,
     1987; A.A.El-Zant, V.G.Gurzadyan, Physica, D, 122, 241, 1998.
 
\end{document}